\documentclass[10pt]{olplainarticle}
\usepackage{float}
\usepackage[hidelinks]{hyperref}
\usepackage{multirow}
\usepackage{tikz}
\title{\centering The influence of Political trust on the acceptance of Violence}

\author[1]{Mario Villagrán}
\affil[1]{Instituto de Sociología Pontificia Universidad Católica de Chile}


\begin{abstract}

The choice of protest tactics in a social movement has often been analyzed based on the demands, participants, and internal characteristics of the movement. However, recent evidence highlights the context or setting in which the demonstration takes place as another key element in the process; Using structural equation modeling, studies have shown a link between high perceptions of injustice in the treatment received by authorities and a greater acceptance of non-normative and/or violent methods of protest. In line with this approach, this article aims to examine the extent to which another form of authority legitimacy --- such as political trust --- affects the overall justification for the use of violence by both protesters and the police. Using longitudinal data from Chile (2016 -- 2019), which captures the collective protests of the ``Social Outbreak'', three analytical approaches --- fixed effects, cross-lagged, and multilevel models --- demonstrate that declining political trust not only weakened public acceptance of police violence but also increased tolerance toward protesters' use of violent tactics. This relationship adds a new dimension to the analysis of violent protests, suggesting that low political trust in many modern states may be a contributing factor to the increasing radicalization of demonstrations in recent years.
\end{abstract}

\begin{document}

\flushbottom
\maketitle
\thispagestyle{empty}

\section*{\centering Introduction}

Until now, the selection of protest strategies in demonstrations has garnered some interest in the literature on social movements. According to \citet{medel2016}, four factors are important in determining protest tactics: the entity toward which the demonstration is directed, the characteristics of the protesting groups, the presence or absence of formal organizations, and the number of participants (p. 168).

Their findings for the period from 2000 to 2012 establish that in Chile, the lower political capital of participants and their limited integration into the productive sphere encourage the adoption of violent demonstration strategies only when their demands require significant institutional reforms to be met (p. 191). However, protests against the state do not typically employ violence; instead, they favor peaceful, legal, and relatively orderly tactics (p. 187).

A turning point in this trend occurred in October 2019 in Chile. Initially triggered by students protesting against an increase in subway fares, the violence quickly spread to forty cities across the country, prompting President Piñera to deploy the military to the streets, ban public gatherings, and impose nighttime curfews—an action that had not been taken by a Chilean president in more than 20 years.

The wave of violent social mobilizations between October 18 and December 2019 resulted in over nine thousand serious incidents \citep{comision2020}. These periods of crisis and protests against institutionalism are not unique to Chile. Similar outbreaks of violence occurred in Argentina in 2001, South Korea in 2017, Ecuador in 2019, Brazil in 2019, and Mexico in 2019.

Recent studies have found a connection between the legitimacy of the political system and the increased acceptance of acts of violence. \citet{gerber2021}, drawing from the social psychology of social movements as well as theories of social justice and legitimacy, examined the effect of perceived unfair treatment by authorities on both normative collective action—such as signing petitions or participating in authorized and peaceful protests—and non-normative action—participating in violent protests or setting up barricades. However, once the effects of legitimacy and its control variables were accounted for, unfair treatment was found to have a direct effect only on normative collective action. In contrast, belief in the legitimacy of authorities significantly and directly influenced both types of action.

Since \citet{pipa1999} studies, there has been a consensus on the decline in support for democratic institutions, or, as \citet{vander2017} describes it, a decrease in political trust in the government, parliament, and political parties. Likewise, a weakened state may struggle to legitimize its monopoly on violence, potentially allowing other groups to face less resistance when resorting to violence.

Based on the above, we propose two central assumptions:

Political and institutional trust serves as a precursor to the state's levels of legitimacy (Easton, 1965, cited \citet{morales2020}; \citet{dammert2014}) and influences citizens' willingness to engage in both traditional and non-traditional channels of participation (protests) (\citet{huet2013}, p. 31; \citet{rivera2019}, p. 574; \citet{gerber2021}).

The use of physical violence tactics in an anti-state demonstration may be facilitated by low political trust in the state. Individuals determine which tactics are "acceptable"—whether the violent repertoire of the protesters or the state's repressive violence—using this trust as one of their criteria (\citet{martinez2016}, p. 29; \citet{gerber2021}, pp. 98–99).

If individuals' levels of political trust influence their overall tolerance for certain displays of physical violence, this could represent another factor affecting the use of such strategies within a cycle of mobilizations. Hence, analyzing this relationship is crucial to understanding the determinants of protest strategy selection.

\section*{\centering Bibliographic Discussion} 

The study of violence has predominantly been approached from two perspectives: as a violation of norms and as the use of excessive force (\citet{bufacci2005}, p. 197). Regarding the use of violence, \citet{martinez2016} identifies two key processes: (1) the denial or distancing between the self and the other , since exercising violence requires reducing the empathy of the perpetrator toward the victim (p. 26); and (2) the valorization of violence, meaning the assessment of its legitimacy based on who is involved, against whom it is directed, and who makes the judgment. Socializing a specific interpretation of the violent act is crucial for attracting potential allies (\citet{martinez2016}, p. 29). In this context, the State can only claim the exclusive and legitimate right to physical force by demonstrating that its use is reasonable (\citet{gerber2016}, pp. 3–4).

At the national level, at least two studies have examined how the public understands violence: the 2015 survey by Diego Portales University (UDP) and the Conflict Module  of the Chilean Longitudinal Social Survey  (ELSOC), conducted in 2016.

The first study examined attitudes toward three types of actions: lynchings  carried out by citizens, excessive use of force by the police  , and the imposition of severe legal penalties (\citet{puga2016}, p. 3). In their analysis, the authors found that the public supports violence against individuals who have committed crimes (\citet{puga2016}, pp. 10–11). Furthermore, this support was strongly influenced by the political and social identification of the respondents (pp. 6 and 11). Social classes was also key in the ELSOC study. According to its findings, individuals from higher social classes are more supportive of police violence against protesters, those from the middle class are more likely to endorse lynchings, and individuals from lower social classes are more likely to justify students cause riots (\citet{gerber2017}, p. 10).   

\subsection*{Conventional and Unconventional Political Participation}

There is currently a consensus that conventional political participation has declined, giving way to new strategies of political action, such as social movements, boycotts, and street protests (\citet{Barnes1979}; \citet{inglehart1997}; \citet{pippa2002}; Inglehart and Catterberg, 2002, as cited in \citet{rivera2019}, p. 559). However, this rise in unconventional political participation seemingly excludes political violence (Fuchs, 1990, as cited in \citet{van2001}). Additionally, regarding success rates and after reviewing 285 campaigns from the 20th century, \citet{ackerman2008} found that non-violent campaigns were twice as successful. Unlike violent ones, they do not contribute to public order narratives \citep{wasow2017} or facilitate state repression (\citet{munoz2019}, p. 4).

The literature suggests that people may resort to violent strategies when riots are framed as forms of political action, particularly when other alternatives are either unavailable or ineffective (Sears and McConahay, 1973, as cited in \citet{enos2019}, p. 12). This is especially true in weakly institutionalized environments, where exercising influence through traditional channels is uncommon (\citet{huet2013}, p. 31). This statement is supported by data from the “Social Outbreak.” For instance, those who participated in protests more than once belonged to the group that held the most critical view of democracy: 65 \% of them believed it was functioning poorly or very poorly. In contrast, non-mobilized opposition of the protests and respondents without a clear stance were far less critical (\citet{cox2021}, p. 22). Furthermore, when it came to justifying the use of illegal or violent tactics to achieve social change, protest supporters and participants were the most likely to endorse them.

In their analysis of the 1992 riots in the United States, \citet{enos2019} add another perspective to the discussion. The authors argue that, although there is a general position disapproving of violent demonstrations, local opinion may be more tolerant due to shared identity, deeper awareness of local circumstances, or fear of new protests (p. 1013).

A second study that more explicitly links the legitimacy of authorities to attitudes toward protests is the 2021 research by \citet{gerber2021} Using data from 2013, the authors found that support for normative collective actions—such as signing petitions or participating in authorized, peaceful protests—grew as the legitimacy of authorities increased. Conversely, support for non-normative actions—such as engaging in violent protests or erecting barricades—declined under the same conditions (pp. 98–99).

At present, there is little doubt today about the sustained decline in political trust \citep{pipa1999}. In this context, Latin America exhibits even lower levels of trust than other regions \citep{dammert2014}. As for Chile, \citet{bargsted2022} point out that several factors undermine both political and general trust: high levels of income inequality, deep class and ethnic divisions, and insufficient social policies that fail to address these issues (p. 2). Moreover, in 2015, multiple cases of corruption and influence peddling became public knowledge, further eroding trust in Chile’s key democratic institutions (\citet{gamboa2016}, p. 126). 

With regard to social mobilizations, \citet{medel2016} highlight an increase in the volume of protests and social conflict since 2006 (p. 165). In this context, concerning the Social Outbreak, COES identifies two factors that may have contributed to its development: (i) a sense of exclusion among the poorest sectors, and (ii) urban segregation in areas with limited access to services and facilities, poor street and housing infrastructure, and unfavorable environmental conditions \citep{prensah2019}.

In a study of mobilized and non-mobilized individuals     during the “Social Outbreak,” \citet{cox2021} found that support for the use of illegal or violent tactics to achieve social change distinguishes supporters and participants from opponents. They also identified key demographic differences between the two groups: the most active protesters were young, with 62 \% under the age of 35, and only 15 \% had not completed high school. In contrast, among non-mobilized opponents, 45 \% had not finished high school, and nearly 40 \% were aged 55 or older (\citet{cox2021}, pp. 10-11).

Thus far, each of the topics discussed provides insight into the factors that shape the selection of protest repertoires. In summary, recent findings suggest that protesters may resort to violent tactics when traditional channels of participation offered by the state are discredited (perceived as lacking legitimacy). Diminished trust in institutions plays a crucial role in determining the public’s tolerance  rejection threshold  toward violent actions, influencing how people judge which forms of violence are deemed legitimate.

Figure 1 illustrates the previous point. The “+” sign for Political Trust and Police Violence Tactics  indicates a positive relationship: as trust increases, so does the justification for police violence. In contrast, the “–” sign for Protesters' Violent Tactics suggests that higher levels of trust correspond to lower acceptance of such tactics. The argument implicitly holds that legitimizing one form of violence diminishes the legitimacy of the other competing alternative.

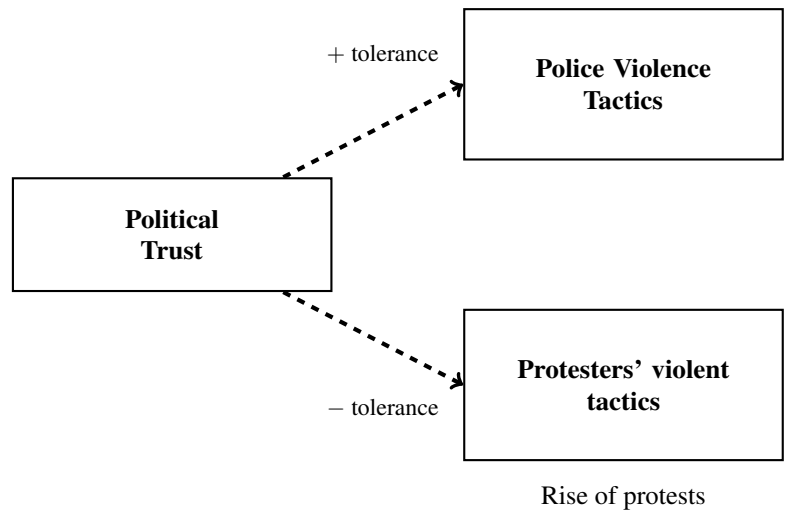
\begin{figure}[H]
    \centering
    \begin{tikzpicture}
    \node[draw, thick, text width=4cm, minimum height=1.5cm, align=center] 
        (1) at (0,0) {\textbf{Political\\ Trust}};
    
    \node[draw, thick, text width=4cm, minimum height=2cm, align=center] 
        (2) at (6,2) {\textbf{Police Violence\\ Tactics}};
    
    \node (t1) at (2.8,2.4) {\small $+$ tolerance};
    
    \node[draw, thick, text width=4cm, minimum height=2cm, align=center] 
        (3) at (6,-2) {\textbf{Protesters' violent\\ tactics}};
    
    \node (4) at (6, -3.5) {Rise of protests};
    
    \node (t2) at (2.8,-2.3) {\small $-$ tolerance};

    \draw[thick, ->, ultra thick, dashed] (1) -- (2.west);
    \draw[thick, ->, ultra thick, dashed] (1) -- (3.west);
    \end{tikzpicture}
    \caption{Proposed relationship between the variables}
    \label{fig1}
\end{figure}

\section*{\centering Methodology} 

The analysis draws on panel data from four waves  collected between 2016 and 2019 as part of the Chilean Longitudinal Social Survey (ELSOC). The respondents were men and women aged 18 to 75 from urban areas, selected through a stratified multi-stage cluster sampling design. A second data source comes from the political activism records in Chile from 2016 to 2019, compiled by the COES Conflict Observatory  using national press reports. This dataset includes several variables, such as the location of the mobilization, commune\footnote{It is the smallest administrative division of Chile and can cover cities, towns, villages and rural areas.}, the start and end dates, the use of violent repertoires, and others.

\subsection*{Study Variables}
\vspace{0.5em}
Dependent variable\\

\textit{Tolerance towards certain forms of violence}.

\vspace{0.5em}

The dependent variables encompass the disposition toward violence in the context of a protest. The ELSOC Survey examined these views under two specific circumstances:

\vspace{0.5em}

\textbf{Question}: “To what extent do you think the following situations are justified or not?”.

\textbf{Situation 1}: The police using force to suppress a peaceful demonstration.

\textbf{Situation 2}: Students throwing stones at the police during a march for educational reform in the country.

\vspace{0.5em}

The statements were answered using a 5-point Likert scale , ranging from "Never justified" to "Always justified". However, because the responses are heavily skewed  towards lower levels of justification, recoding this variable into a dichotomous format is going to be used. This will retain the "Never justified" response while grouping all other responses that indicate some acceptance of violence into a single category labeled "Could be justified." This approach provides a more equitable distribution among the positions. 

\vspace{0.5em}

Independent variables\\

\textit{Political Trust Index }.

\vspace{0.5em}

The main independent variable for the purposes of this study. :
It was prepared from the item that ELSOC used to inquire about the level of trust of the respondents towards various institutions or government figures.

The question was answered using a 5-point scale ranging from “none” to “full confidence” in the Government, political parties, the National Congress, the Judiciary, the President of the Republic and the police.

The criteria for constructing the Political Confidence Index from this item used two procedures:

1) A factor analysis, which would support the consistency of the behavior of the factor loadings, in the 4 waves of the survey. That is, that the components grouped the same institutions and actors in each wave of application.

2) A Cronbach's Alpha over 0.7 of the proposed index, for each wave of application.

The results of the proposed index are described in table no. 1:

\begin{table}[ht]
\centering
\small 
\begingroup
\renewcommand{\arraystretch}{1.4} 
\resizebox{\textwidth}{!}{%
\begin{tabular}{|c|c|cccc|}
\hline
\multirow{2}{*}{\textbf{Components}} & \multirow{2}{*}{\textbf{Items}} & \multicolumn{4}{c|}{\textbf{Cronbach's Alpha}} \\ \cline{3-6}
                                     &                                & \textbf{Wave 1} & \textbf{Wave 2} & \textbf{Wave 3} & \textbf{Wave 4} \\ \hline
\multirow{2}{*}{1} 
  & Government                  & \multirow{4}{*}{0.79} & \multirow{4}{*}{0.82} & \multirow{4}{*}{0.81} & \multirow{4}{*}{0.80} \\
  & President of the Republic &                        &                        &                        &                        \\ \cline{1-2}
\multirow{3}{*}{2} 
  & National Congress         &                        &                        &                        &                        \\
  & Judiciary                 &                        &                        &                        &                        \\
  & Political parties         &                        &                        &                        &                        \\ \hline
\end{tabular}
}
\endgroup
\caption{Cronbach's Alpha by Component and Item}
\label{tab:cronbach_centered}
\end{table}

Both empirically and thematically, the item that asks about trust in the police is not associated with other actors and political institutions. This result is consistent with what \citet{morales2020} found in a similar application. Finally, the political confidence index in this study includes the evaluation of the Government, political parties, the National Congress, the Judiciary and the President of the Republic.

\vspace{1em}

Control variables\\

In the analysis, a series of control variables were utilized, including gender, age, political orientation, educational level, subjective social class, participation in marches and mobilizations, wave, and the type of activism recorded in the respondent's commune. The aforementioned variable was constructed using data from the political activism database and considers four possible situations:

1)	Communes  where there is no record of any political activism during the first two months of collecting responses from ELSOC (the same time frame applies to each of the following alternatives). 

2)	Communes where there are records of the initiation of demonstrations or protests that did not involve violence.

3)	Communes where there is evidence of the start of a demonstration or protest that did involve violence (most often occurring alongside protests that did not, according to the data). 

4)	Communes with information about other forms of political activism that are not necessarily identified with protests or marches, such as “hunger strikes,” “attacks on bystanders or uninvolved third parties,” and “hostage-taking or kidnappings.”

\subsection*{Types of analisys}
\vspace{0.5em}
The first phase consists of assessing the association between the level of trust among subjects and changes in their level of tolerance, using a fixed effects model, which, in formal terms, is as follows:

\begin{equation} \label{eq:model} 
y_{it} = \beta_0 + \beta_1 x_{1it}+ \beta_2 x_{2it}+\mu_i+e_{it}
\end{equation}

Here, $y_{it}$ represents the degree of tolerance for the use of a particular violence tactic, for example, tolerance for violence used by protesters. Similarly, $x_{1it}$ in $\beta_1$ represents the individual's political trust index, while $x_{2it}$ in $\beta_2$ corresponds to a vector that includes all the control variables indicated above.

One advantage of this estimation method for panel data analysis is that it helps control for any omitted variable bias resulting from unobserved characteristics invariant over time and not explicitly included in the model. Likewise, serial correlation issues common in time series data can be addressed by using clustered standard errors at the respondent level. However, fixed effects models are sensitive to endogeneity problems, which are likely present in the data. Therefore, it is necessary to complement the results with an alternative method of assessment to ensure a more robust overall result.

According to the literature, one way to address endogeneity problems is by using a cross-lagged model, which, as \citet{kearney2017} states, analyzes the causal influences between two variables, providing evidence on the direction of causality. Some scholars have used this approach specifically for this purpose (\citet{sonder2016}; \citet{eveland2005}).

In this second phase, we aim to determine which variable has a greater influence on the other: political trust on tolerance, or tolerance on political trust. The equations addressing this question are:

\begin{equation} \label{eq:model2}
y_{1it} = \beta_0 + \rho y_{1it-1} + \beta_1 x_{2it-1} + \beta_2 x_{3it-1} + e_{it}
\end{equation}

\begin{equation} \label{eq:model3}
x_{2it} = \beta_0 + \beta_1 y_{1it-1} + \rho x_{2it-1} + \beta_2 x_{3it-1} + e_{it}
\end{equation}

Here, $Y_{1it}$ corresponds to the level of tolerance for a type of violence, for example, police repression. Meanwhile, the independent variables, are introduced into the model using the values recorded in the wave preceding $Y_{it}$, hence the subscript $-1$.

Cross-lagged models include a lagged dependent variable as a predictor (represented by the coefficient $\rho Y_{1it-1}$), which allows us to infer what portion of the variance, unexplained by the variable itself, is effectively associated with the other analyzed conditions.

In turn, $x_{2it-1}$ in $\beta_1$ represents the individual's political trust index. Similarly, $x_{3it-1}$ in $\beta_3$ describes a vector that contains all control variables.

Subsequently, a mirror equation of~\eqref{eq:model2} is estimated, formally specified in Equation~\eqref{eq:model3}, where trust is now placed as the dependent variable and tolerance from the previous wave as the independent variable. The comparison of the cross-lagged coefficients between both models will allow us to gather evidence on the stronger direction of effect.

Due to the distribution of tolerance responses being strongly skewed toward rejection positions, not all response options of the variable provide a sufficient number of selections to allow for proper modeling. For this reason, the response options were recoded into only two categories: ``It is never justified'' and ``It could be justified.'' A multilevel logistic model nested at the individual level was then employed to reassess the relationship between political trust and tolerance.

Thus, in formal terms, the proposed model is as follows:

\begin{equation} \label{eq:model4}
\log \left[ \frac{\Pr(y_{ij} = 1)}{1 - \Pr(y_{ij} = 1)} \right] = 
\gamma_{00} + \gamma_{01} \bar{x}_{1j} + \gamma_{02} (x_{1ij} - \bar{x}_{1j}) + \gamma_{03} \textbf{control} + \mu_j + e_{ij}
\end{equation}

The dependent variable corresponds to the logarithm of the odds ratio between the probability of selecting the ``It could be justified'' option (\( y_{ij} = 1 \)) and the alternative representing total rejection of violence.

Regarding the independent variables: \( x_{1ij} \) reflects the individual's political trust values, decomposed into their respective between-individual and within-individual variances. Here, \( \bar{x}_{1j} \) denotes the person's average trust level over the entire period, while \( x_{1ij} - \bar{x}_{1j} \) represents the individual’s annual trust values centered on their mean (within-individual differences).

The ``Control'' variable represents a vector that includes all control variables used in the study. \( \mu_j \) is the random effect that accounts for deviations in the odds ratio across subjects over all waves, based on \( \gamma_{00} \) (the grand mean). Meanwhile, \( e_{ij} \) describes the deviation of a person’s annual observations from the grand mean.

This approach not only verifies the influence between variables but also quantifies the effect of political trust changes on the probability of tolerating or rejecting violence.

\section*{\centering Results}
\subsection*{Fixed effects models}
\vspace{0.5em}

Table~\ref{tab:fixed_effects_violence} presents two types of models. The first model identifies the justification of police violence as the dependent variable, while the second model focuses on the acceptance of violent repertoires by protesters. The results reveal a significant association between political trust and the justification of violence variables, with a 99\% confidence level for repression in Model~1 and a 99.9\% confidence level for the case of protesters in Model~2. According to the coefficients, for each 1-point increase in the political trust scale, the justification of police actions rises by 0.066 on the variable scale, while the acceptance of violence by protesters decreases by 0.065, controlling for the other variables in the model. These results confirm that an increase in an individual's political trust reinforces their tolerance for police repression while diminishing their acceptance of the use of violent repertoires by protesters.

In terms of attendance at marches, the second model indicates a positive effect of participation on the justification of violence by protesters. This result is consistent with the findings of \citet{cox2021} who observed a baseline tendency among individuals who protested or supported the mobilizations during the social outbreak to endorse the use of violent repertoires in the protests. On the other hand, the emergence of violent activism within the respondent's commune is positively correlated with the justification of violent repertoires, a relationship previously anticipated by \citet{enos2019}.

\begin{table}[H]
\centering
\caption{Fixed Effects Models: Attitude towards Violence}
\label{tab:fixed_effects_violence}
\renewcommand{\arraystretch}{1.2}
\begin{tabular}{lcc}
\toprule
 & \textbf{Model 1} & \textbf{Model 2} \\
\midrule
Political trust              & 0.066**     & $-0.065^{***}$ \\
                             & (0.021)     & (0.018)        \\
Protest attendance           & 0.021       & 0.293***       \\
                             & (0.032)     & (0.033)        \\
Close disruptive activism    & $-0.032$    & 0.033          \\
                             & (0.030)     & (0.025)        \\
Close violent activism       & $-0.079$    & 0.148***       \\
                             & (0.044)     & (0.038)        \\
Other forms of activism      & 0.031       & 0.248***       \\
                             & (0.065)     & (0.056)        \\
Wave 2                       & $-0.097^{**}$ & $-0.098^{***}$ \\
                             & (0.030)     & (0.023)        \\
Wave 3                       & $-0.032$    & $-0.111^{***}$ \\
                             & (0.042)     & (0.029)        \\
Wave 4                       & $-0.265^{***}$ & $-0.034$     \\
                             & (0.058)     & (0.043)        \\
\midrule
$R^2$                        & 0.035       & 0.044          \\
Adj. $R^2$                   & $-0.519$    & $-0.503$       \\
Num. obs.                    & 12062       & 12077          \\
\bottomrule
\end{tabular}

\vspace{0.5em}
\begin{flushleft}
\footnotesize
\textit{Note:} *** $p < 0.001$, ** $p < 0.01$, * $p < 0.05$. \\
\vspace{0.5em}
\textit{Source:} Own elaboration based on ELSOC 2016--2019 and the Observatory of Conflicts.
\end{flushleft}
\end{table}

\subsection*{Interactions in   Fixed effects models}
\vspace{0.5em}

According to the results, those who mobilize are more likely to accept violence as a valid protest strategy. Therefore, analyzing the role of institutional trust provides insight into the underlying mechanisms driving this inclination.

In the following table (Table~\ref{tab:interaction_models}), Models~3 and~4 include an interaction between political trust and participation in marches to examine how these variables behave together. According to the results from Model~3, where the dependent variable is police violence, the interaction coefficient is not significant. Therefore, the effect of political trust does not vary considerably across different levels of participation. However, when positioning the violence of protesters as the dependent variable (Model~4), we find an even greater deterrent effect of political trust among those who attend marches. This indicates that their baseline inclination towards using this protest tactic (main effect coefficient) can be entirely reversed depending on the presence of sufficient levels of political trust. These results align with certain findings in the literature, suggesting that individuals with low levels of political trust are most inclined to favor violent activism.

Models~5 and~6 analyze the interaction between political trust and the onset of protests in the respondent's commune. In the case where the dependent variable is police repression, as shown in Model~5, the effect of trust does not vary significantly across different types of activism. However, the scenario shifts when the dependent variable is protester violence. The results of Model~6 suggest that higher levels of political trust lead to stronger disapproval of this form of protest, particularly in areas where such activism takes place.

\begin{table}[H]
\centering
\caption{Fixed Effects Models: Interactions with Participation and Activism Type}
\label{tab:interaction_models}
\renewcommand{\arraystretch}{1.2}
\begin{tabular}{lcccc}
\toprule
 & \textbf{Model 3} & \textbf{Model 4} & \textbf{Model 5} & \textbf{Model 6} \\
\midrule
Political trust                      & 0.063**    & $-0.042^{*}$ & 0.030     & $-0.054^{*}$ \\
                                     & (0.022)    & (0.018)      & (0.030)   & (0.024)      \\
Protest attendance                   & $-0.014$   & 0.544***     & 0.021     & 0.287***     \\
                                     & (0.075)    & (0.082)      & (0.032)   & (0.033)      \\
Close disruptive activism            & $-0.032$   & 0.032        & $-0.163^{*}$ & 0.005     \\
                                     & (0.030)    & (0.025)      & (0.078)   & (0.062)      \\
Close violent activism               & $-0.079$   & 0.147***     & $-0.179^{*}$ & 0.281***  \\
                                     & (0.044)    & (0.038)      & (0.085)   & (0.076)      \\
Other forms of activism              & 0.031      & 0.247***     & 0.071     & 0.240        \\
                                     & (0.065)    & (0.056)      & (0.186)   & (0.151)      \\
Wave 2                               & $-0.097^{**}$ & $-0.098^{***}$ & $-0.095^{**}$ & $-0.098^{***}$ \\
                                     & (0.030)    & (0.023)      & (0.030)   & (0.022)      \\
Wave 3                               & $-0.032$   & $-0.114^{***}$ & $-0.030$  & $-0.115^{***}$ \\
                                     & (0.042)    & (0.029)      & (0.042)   & (0.029)      \\
Wave 4                               & $-0.264^{***}$ & $-0.042$     & $-0.259^{***}$ & $-0.039$     \\
                                     & (0.059)    & (0.043)      & (0.059)   & (0.043)      \\
Pol.trust*Protest attendance         & 0.020      & $-0.145^{***}$ &         &             \\
                                     & (0.041)    & (0.042)      &          &             \\
Pol.trust*Disruptive activism        &            &              & 0.072    & 0.019        \\
                                     &            &              & (0.040)  & (0.029)      \\
Pol.trust*Violent activism           &            &              & 0.058    & $-0.078^{*}$ \\
                                     &            &              & (0.044)  & (0.037)      \\
Pol.trust*Other activism             &            &              & $-0.019$ & 0.008        \\
                                     &            &              & (0.094)  & (0.072)      \\
\midrule
$R^2$                                & 0.035      & 0.047        & 0.035    & 0.046        \\
Adjusted $R^2$                       & $-0.519$   & $-0.500$     & $-0.519$ & $-0.502$     \\
Num. obs.                            & 12062      & 12077        & 12082    & 12097        \\
\bottomrule
\end{tabular}

\vspace{0.5em}
\begin{flushleft}
\footnotesize
\textit{Note:} *** $p < 0.001$, ** $p < 0.01$, * $p < 0.05$.\\
\vspace{0.5em}
\textit{Source:} Own elaboration based on ELSOC 2016--2019 and the Observatory of Conflicts.
\end{flushleft}
\end{table}

\subsection*{Cross-lagged models}
\vspace{0.5em}

In this section, the results will be presented through diagrams illustrating the behavior of key variables across different waves.
\vspace{0.3em}
\begin{figure}[ht]
    \centering
    \includegraphics[width=\textwidth]{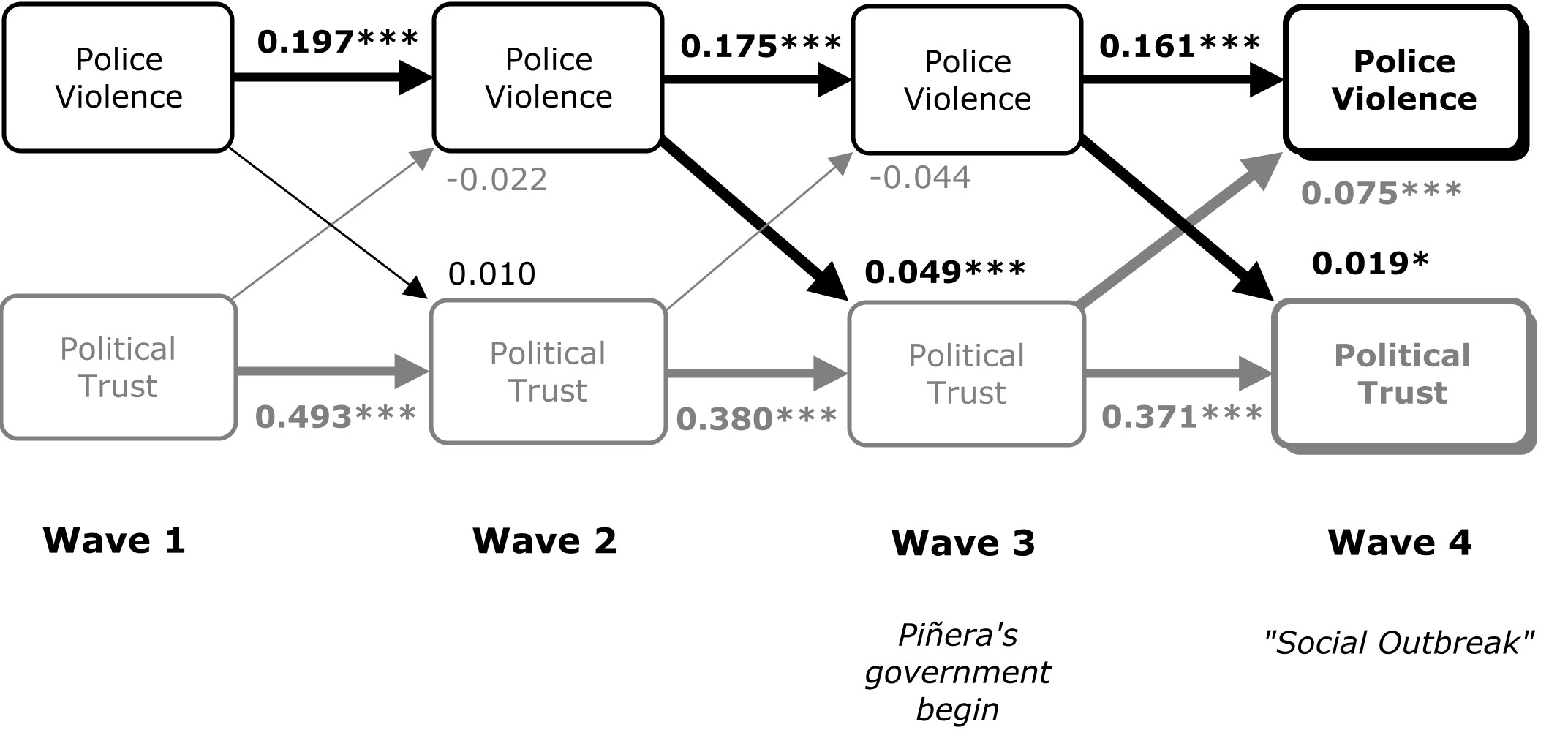}
    \caption{Relationship between Police Violence and Political Trust across the Waves}
    \label{fig:crosslagged_political_trust_police}
    \vspace{0.5em}
\footnotesize\textit{Source:} Own elaboration based on ELSOC 2016--2019 and the Observatory of Conflicts.
\end{figure}

Figure~\ref{fig:crosslagged_political_trust_police} illustrates the relationship between levels of political trust and tolerance for police repression. During the “Social Outbreak” and prior to it, there is a significant positive effect of acceptance of violence from the previous wave on political trust in the subsequent wave. However, as the crisis develops, the latest wave reveals a significant relationship in the opposite direction, consistent with the findings from the fixed-effects models. Specifically, the higher levels of political trust recorded in Wave~3 correspond to increased tolerance for the use of violence in Wave~4, thereby weakening the influence of justifications for police actions on future political credibility.

\begin{figure}[H]
    \centering
    \includegraphics[width=\textwidth]{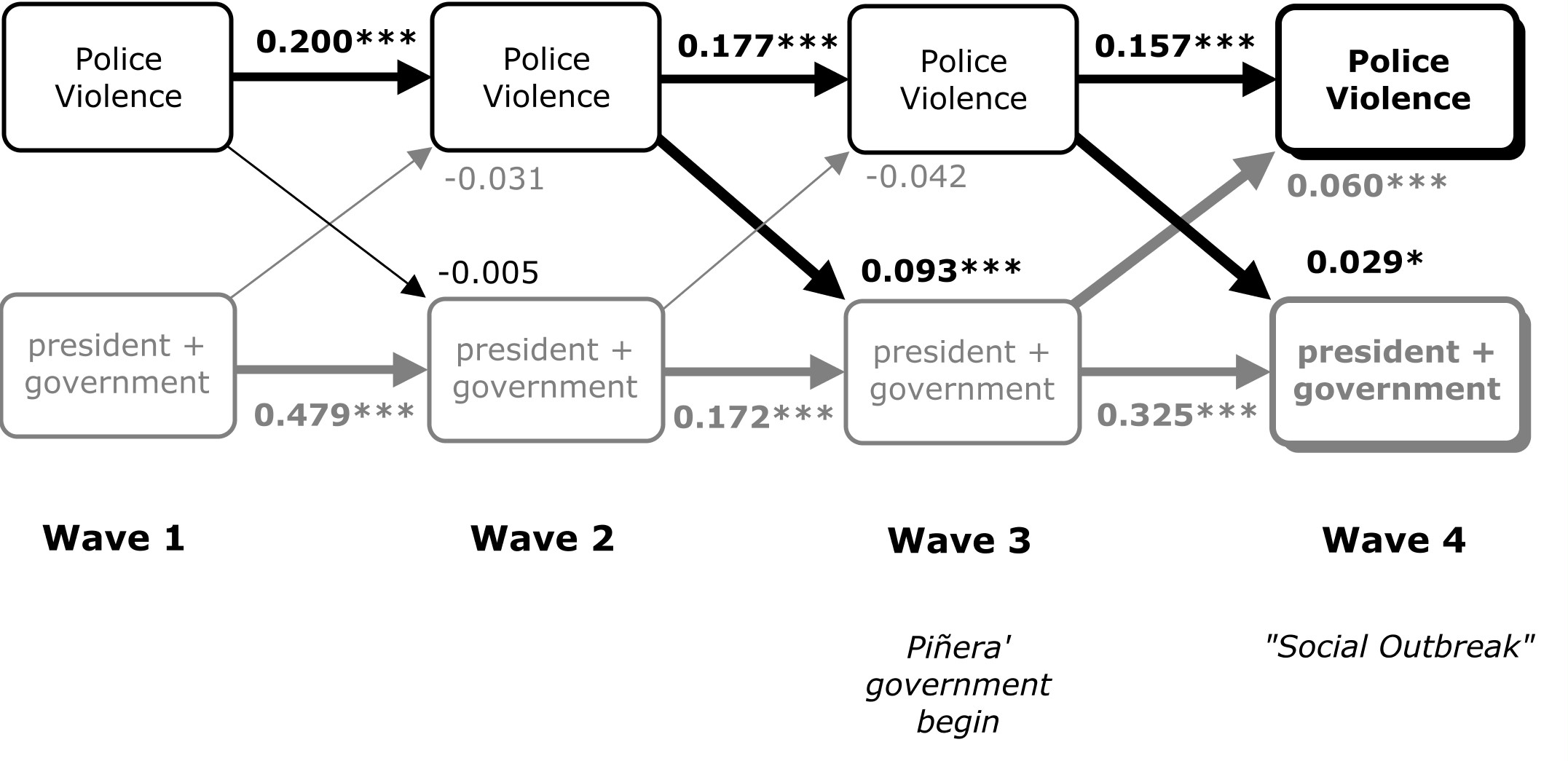}
    \caption{Relationship between Protester Violence  and Political Trust across Waves}
    \label{fig:trust_police_componente1}
    \vspace{0.5em}
\footnotesize\textit{Source:} Own elaboration based on ELSOC 2016--2019 and the Observatory of Conflicts.
\end{figure}

Figure~\ref{fig:trust_police_componente1} describes the influence between political trust and the justification of violence by protesters. According to the results, there is no significant relationship between these two variables outside the crisis period. It is only after the emergence of the “Social Outbreak” that an effect appears from prior levels of trust toward subsequent tolerance for the use of violent repertoires. This influence shows the highest coefficient reported in the cross-lagged models, exerting an impact similar to that of the respondents' previous positions on violence. Lastly, it is worth noting that, consistent with the findings from the fixed-effects models, the relationship between the variables is negative: as levels of trust increase, tolerance for the use of violent repertoires decreases.
\vspace{0.5em}
\subsection*{Multilevel Logistic Models  and Calculation of Predicted Probabilities}
\vspace{0.5em}
As shown in Table~\ref{tab:multilevel_model}, the results for the dependent variables ---police violence and protester violence--- are estimated using a binary recoding of tolerance, where 0 indicates ``Never justified'' and 1 means ``Could be justified.'' The coefficient for \textit{Centered Political Trust}, which captures the within-individual variance, is statistically significant at the 99\% confidence level. This result, presented in Model 7, indicates that individuals with higher levels of political trust are significantly more likely to accept some level of police repression, even after accounting for all other covariates in the model.

\begin{table}[H]
\centering
\caption{Multilevel Models: Position on Police Violence}
\label{tab:multilevel_model}
\renewcommand{\arraystretch}{1.2}
\begin{tabular}{lc}
\toprule
 & \textbf{Model 7} \\
\midrule
Intercept                                & $-1.034^{***}$ \\
                                        & (0.209)         \\
Centered Political Trust                & $0.178^{***}$   \\
                                        & (0.049)         \\
Mean Political Trust                    & $0.380^{***}$   \\
                                        & (0.054)         \\
Protest Attendance                      & $-0.197^{**}$   \\
                                        & (0.074)         \\
Close Disruptive Activism              & $-0.062$        \\
                                        & (0.060)         \\
Close Violent Activism                 & $-0.134$        \\
                                        & (0.078)         \\
Other Forms of Activism                & 0.154           \\
                                        & (0.120)         \\
Wave 2                                  & $-0.218^{**}$   \\
                                        & (0.067)         \\
Wave 3                                  & $-0.052$        \\
                                        & (0.065)         \\
Wave 4                                  & $-0.738^{***}$  \\
                                        & (0.076)         \\
\midrule
Sigma$^2$ (Intercept)                   & 0.975           \\
AIC                                     & 13549.269       \\
BIC                                     & 13680.189       \\
Log Likelihood                          & $-6753.635$     \\
Num. Observations                       & 11449           \\
Num. Respondents                        & 4417            \\
\bottomrule
\end{tabular}

\vspace{0.5em}
\begin{flushleft}
\footnotesize
\textit{Note:} *** $p < 0.001$, ** $p < 0.01$, * $p < 0.05$.\\
\vspace{0.5em}
\textit{Source:} Own elaboration based on ELSOC 2016--2019 and the Observatory of Conflicts.
\end{flushleft}
\end{table}

\vspace{1em}

In the case of protester violence in Model 8 (Table ~\ref{tab:multilevel_model_protesters}), the within-variance of political trust indicates that changes within individuals significantly influence their stance toward this type of repertoire, with a significance level of 99.99\%. Rejection of violent activism increases as political trust rises. This behavior aligns with \citet{martinez2016}, who argues that observers consider and evaluate the characteristics of actors competing to employ violence, before deciding whom to support.

\begin{table}[H]
\centering
\caption{Multilevel Models: Position on Protester Violence}
\label{tab:multilevel_model_protesters}
\renewcommand{\arraystretch}{1.2}
\begin{tabular}{lc}
\toprule
 & \textbf{Model 8} \\
\midrule
Intercept                                & $-0.018$         \\
                                         & (0.252)          \\
Centered Political Trust                 & $-0.253^{**}$    \\
                                         & (0.064)          \\
Mean Political Trust                     & $-0.177^{**}$    \\
                                         & (0.064)          \\
Protest Attendance                       & $1.015^{***}$    \\
                                         & (0.077)          \\
Close Disruptive Activism                & $-0.154^{*}$     \\
                                         & (0.076)          \\
Close Violent Activism                   & 0.097            \\
                                         & (0.090)          \\
Other Forms of Activism                  & $0.477^{***}$    \\
                                         & (0.143)          \\
Wave 2                                   & $-0.253^{**}$    \\
                                         & (0.089)          \\
Wave 3                                   & $-0.254^{**}$    \\
                                         & (0.085)          \\
Wave 4                                   & $0.391^{***}$    \\
                                         & (0.087)          \\
\midrule
Sigma$^2$ (Intercept)                    & 0.979            \\
AIC                                      & 9735.115         \\
BIC                                      & 9866.035         \\
Log Likelihood                           & $-4846.557$      \\
Num. Observations                        & 11461            \\
Num. Respondents                         & 4417             \\
\bottomrule
\end{tabular}

\vspace{0.5em}
\begin{flushleft}
\footnotesize
\textit{Note:} *** $p < 0.001$, ** $p < 0.01$, * $p < 0.05$. \\
\vspace{0.5em}
\textit{Source:} Own elaboration based on ELSOC 2016--2019 and the Observatory of Conflicts.
\end{flushleft}
\end{table}

Another important point to address is the significance of the coefficient for \textit{mean political trust} over the entire period, which captures respondents’ long-term trends throughout the study. As the results show, this variable is statistically significant and has a greater impact on attitudes toward police repression than contingent trust, referred to as \textit{centered political trust}. In contrast, tolerance for violent protest repertoires reveals the opposite effect, with annual trust exerting a stronger influence in shaping individuals' stances.

It is also possible to transform the results into probabilities to evaluate how an ``average'' individual's stance on violence varies according to their levels of political trust. The \textit{centered trust} variable originally ranges from $-2$ to $2$, relative to the respondent’s mean political trust. A one-point change, for example, reflects a shift from ``No trust'' to ``Little trust,'' while a two-point increase represents a more significant transition, such as from ``Little trust'' to ``Considerable trust.''

\begin{figure}[H]
    \centering
    \includegraphics[width=\textwidth]{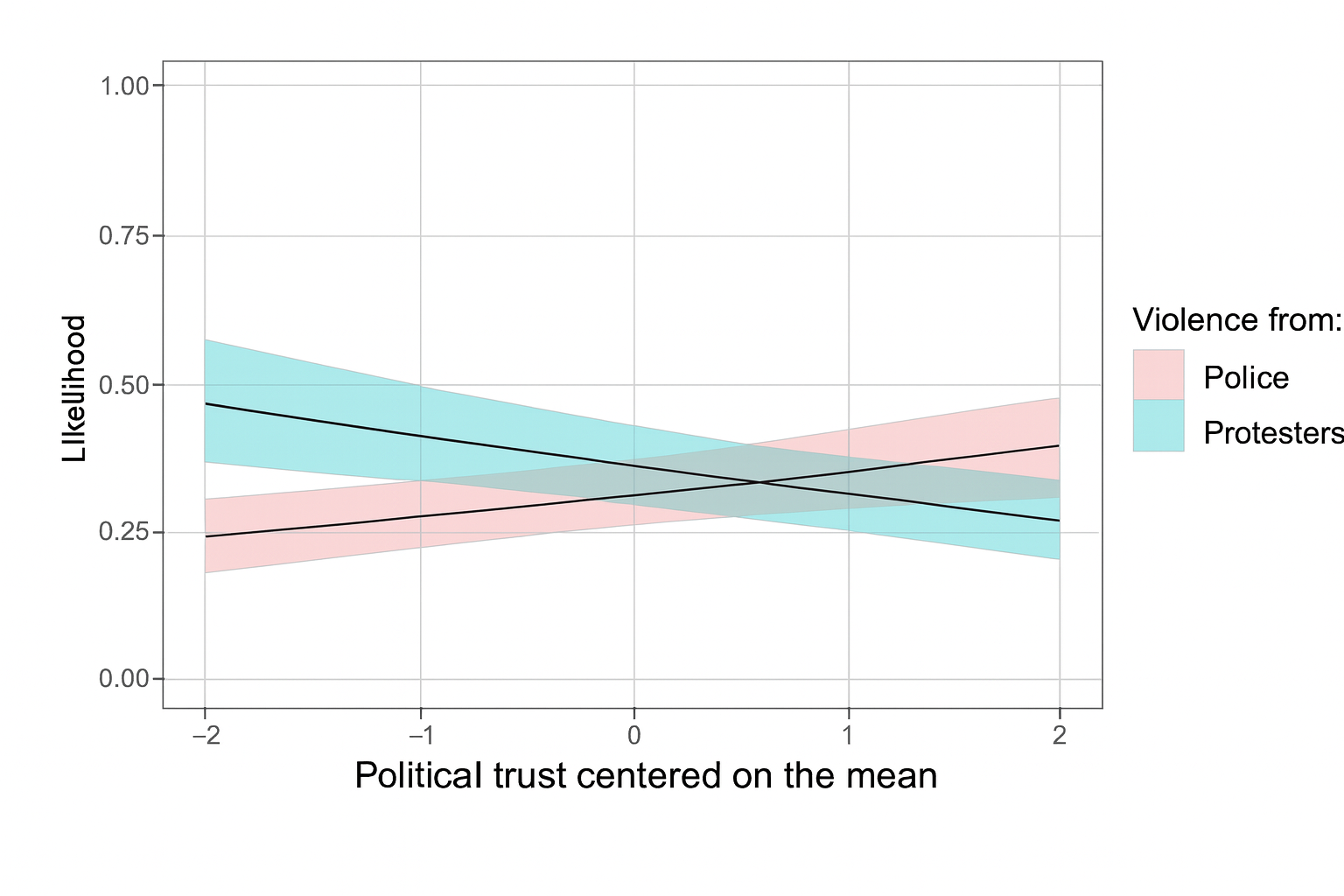}
    \captionsetup{labelformat=empty}
    \caption{\textbf{Graphic 1:} Relationship between Protester Violence and Political Trust across Waves}
    \label{fig:trust_police_componente1}
\vspace{0.5em}
\footnotesize\textit{Source:} Own elaboration based on ELSOC 2016--2019 and the Observatory of Conflicts.
\end{figure}

The results indicate that in scenarios where political trust is most unfavorable, the probability of accepting police repression is slightly higher than that of accepting protester violence, at 25.7\% compared to 25.1\%, respectively.

The threshold for accepting violent repertoires decreases more rapidly with a one- or two-point increase above the average trust level than the corresponding decline in acceptance of police actions resulting from the same change in the opposite direction. Specifically, the probability of accepting violent protest increases from 36\% to 42\% with a one-point negative deviation from the individual’s mean trust level, and to 48\% with a two-point change. In contrast, the likelihood of accepting police repression decreases from an initial 33\% to 29\%, and then to 26\% for the same range of variation. An increase in political trust relative to the respondent’s average reverses the scenario: a one-point or two-point increase enhances the likelihood of validating police violence to 37\% and 41\%, respectively. In contrast, the acceptance of violent protest repertoires drops to 30\% and 25\% for the same trust increments.

\clearpage
\section*{\centering Conclusion}

The objective of this document has been to analyze whether levels of political trust influence the acceptance of violence during protests. In line with this purpose, the results confirm that, both during the \textit{“Social Outbreak”} and in other forms of activism from 2016 to 2019, individuals effectively use their levels of trust in the political system as one of their criteria. Thus, higher levels of trust are associated with a greater likelihood of validating police actions to some degree and rejecting violent protest strategies. Conversely, at minimal levels of trust, the scenario favors the violent actions of protesters over those of the police.

One of the most significant contributions of these results is the demonstration of how the erosion of credibility in the political system serves as a risk factor for the radicalization of protest cycles. In this context, the emergence of unrest transforms political trust into a resource that shapes individuals' positions on violence. Moreover, it is among those who are more inclined toward violent actions---such as activists \citep{cox2021} or those exposed to similar forms of activism within their own communes \citep{saunders2014}---that political trust influences and more significantly reduces this inclination.

The results of \citet{gerber2017} suggested that low legitimacy fostered pro-violence stances during protests. Accordingly, a high level of political trust should reduce this inclination, thereby moderating the initial tendencies of both protesters and observers. Our findings confirm this moderating role; the effect was particularly significant in shaping protesters’ acceptance of violence.

For an average activist, the results demonstrate a tendency to support police repression following a one-point change above their mean level of trust (for example, moving from \textit{``no trust''} to \textit{``little trust''}). Conversely, a one-point decrease in trust increases the likelihood of justifying the use of violent repertoires. This gap becomes more pronounced with two-point changes in trust levels. However, regardless of the magnitude, the validation of one alternative inevitably leads to a greater rejection of the competing form of action.

As the state loses support from its citizens, its legitimacy to use violence diminishes, along with its monopoly over it.

In conclusion, the analysis found a consistent relationship between political trust and the thresholds for accepting violent actions. This result supports \citet{martinez2016} that, within the process of valorization, the characteristics of the actors competing to employ violence are relevant to the observer who decides whom to support.

It also extends the original link between legitimacy and valorization described by \citet{gerber2021} to other forms of institutional credibility, such as political trust.

Therefore, even though factors such as participants’ lower political capital and limited integration into the productive sphere may influence the adoption of violent strategies \citep{medel2016}, political trust functions as a framework through which observers assess the legitimacy of violent actions---especially in contexts where protesters and the state are in direct opposition.

Thus, the erosion of trust poses two challenges: it not only increases citizens’ resistance to the state’s use of force to ``restore public order,'' but also reduces the condemnation of radical repertoires---particularly among activists---thereby facilitating their spread.

The declining trend in credibility in the political system is a significant problem, as evidenced by various consequences that underscore its importance in the current context. This situation highlights the urgent need to address its management and recovery.

\clearpage

\bibliography{sample}

\end{document}